\documentclass[
    ,a4paper
    ,final            
  ]
  {aipproc}

\layoutstyle{6x9}
\usepackage{amsmath, amssymb}

\newcommand{\half}{ \frac{1}{2} }
\newcommand{\thalf}{ \tfrac{1}{2} }

\newcommand{\tquar}{ \tfrac{1}{4} }
\newcommand{\bra}[1]{\left\langle #1 \right|\,}
\newcommand{\ket}[1]{\,\left| #1 \right\rangle}
\newcommand{\expect}[1]{\left\langle {#1} \right\rangle}
\newcommand{\bracket}[2]{\left\langle {#1} \,\vrule\, {#2} \right\rangle}

\newcommand{\set}[1]{\left\{ #1 \right\}}
\newcommand{\goesto}{\rightarrow}

\newcommand{\sub}[1]{_{\text{#1}}}

\newcommand{\abs}[1]{\left\lvert {#1} \right\rvert}
\newcommand{\norm}[1]{\left\lVert {#1} \right\rVert}
\newcommand{\floor}[1]{\left\lfloor #1 \right\rfloor}

\newcommand{\mx}[1]{\mathrm \mathbf{#1}}

\begin{document}

\title{How to calculate correlation functions of Heisenberg chains}


\author{Rob Hagemans}{
  address={Instituut voor Theoretische Fysica, Universiteit van Amsterdam, Amsterdam, the Netherlands}
}

\author{Jean-S\'{e}bastien Caux}{
  address={Instituut voor Theoretische Fysica, Universiteit van Amsterdam, Amsterdam, the Netherlands}
}

\author{Jean Michel Maillet}{
  address={Laboratoire de Physique, \'{E}cole Normale Sup\'{e}rieure de Lyon, Lyon, France}
}

\begin{abstract}
We describe a method for calculating dynamical spin-spin correlation functions in the isotropic and anisotropic antiferromagnetic Heisenberg models. Our method is able to produce results with high accuracy over the full parameter space.
\end{abstract}

\maketitle


\section{Introduction}
One-dimensional quantum systems stand out because the effect of interactions is unusually strong, which turns them into a very suitable arena for the study of strong correlations. Another striking and very useful feature of certain one-dimensional systems is the fact that many of their properties can be calculated exactly, using methods that are unavailable for higher-dimensional systems. For a long time, however, the reach of such calculations has been limited to static properties. Indeed, calculating dynamical properties within the framework provided by exact methods has for a long time been regarded as hard, possibly even intractable. This is changing, however. With the advent of exact expressions for form factors in one-dimensional spin models, the possibility is opened up to calculate dynamical quantities.

One such quantity that is of considerable interest is the {\em dynamical structure factor}. This quantity, defined for a general spin operator $S^a_j$ on the $j$th site of an $N$-site chain as
\begin{align}
S^{a\bar a}(q,\omega) 
	&:= \frac{1}{N}\sum_{j,j'=1}^{N} e^{iq(j-j')} \int_{-\infty}^\infty dt~ e^{i\omega t}\expect{ S^a_j(t)\,S^{\bar a}_{j'}(0) }\sub{c} 
	~,
\end{align}
(where the subscript c indicates we are to take a connected correlator) owes part of its importance to its role in describing inelastic neutron-scattering experiments on quasi-one-dimensional crystals, where it is proportional to the intensity of scattered neutrons at the given momenta and energies. 

Neutron scattering is used to probe the magnetic structure of materials. In a quasi-one-dimensional crystal, the coupling between sites in one direction has a much higher energy scale than in the two others. If one does measurements at an energy scale that is intermediate between these two scales, the system behaves as a one-dimensional system.
In many such materials, the local magnetic moments have two low-lying levels, meaning that the systems can effectively be described as spin-$\thalf$ chains, bringing us into a field of research that has been of theorists' interest for quite a long time: that of Heisenberg chains.

Heisenberg's spin model, nowadays also known as the XXX model, originally consisted of a lattice of spins-$\frac{1}{2}$ with isotropic nearest-neighbour couplings. In one dimension, this model was solved by Hans Bethe, who presented a method\cite{Bethe1931} to determine its eigenfunctions and their energies. A more general spin model was introduced by R. Orbach\cite{Orbach1958}. This model, which is also the one we shall be concerned with in this article, features an anisotropy in the spin-spin interaction in the $z$ direction (and is hence known as the XXZ model), as well as an external field parallel to that anisotropy. It is given by the Hamiltonian
\begin{equation}
\label{eqHeisenbergModel}
H =  J \sum_{j=1}^N S^x_j S^x_{j+1} + S^y_j S^y_{j+1} + \Delta(S^z_j S^z_{j+1} - \tquar) - h S^z_j
~,
\end{equation}
where $J$ is the coupling strength, $\Delta$ is the anisotropy parameter, $h$ the external magnetic field and $S^x_j$, $S^y_j$, $S^z_j$ are the usual spin-$\half$ operators acting on chain site $j$. 

Bethe's method---known as the Bethe Ansatz---has been employed to find static properties of many one-dimensional quantum systems. However, due to the intricate form in which the eigenstates are given by the Bethe Ansatz, calculating such quantities as norms, overlap functions, and correlation functions is not an easy task.

\section{Calculation scheme}
\subsection{Bethe's Equations}
The Hamiltonian \eqref{eqHeisenbergModel} conserves the spin component in the $z$ direction (as well as the total spin). Therefore, the Hilbert space it acts on separates into disjunct subspaces with different expectation values of the magnetisation. Throughout this paper, we shall specify the subspace by the number of spins pointing downward, $M := N/2 - \expect{ \sum_{j=1}^N S^z_j }$. 
Throughout this paper, the {\em reference state} is defined to be the state with all spins pointing upward, $\ket{0} := \bigotimes_{j=1}^N \ket{\uparrow}$. The chain length $N$ is always assumed to be even and the boundary is periodic. If the Hamiltonian is defined as above, the energy of the reference state equals $-J h N/2$. 
From this reference state, all states with $M$ up to $N/2$ can be reached with a Bethe Ansatz; to reach the other states, one should start from a reference state with all spins down.

Bethe's solution gives the un-normalised wave functions $\ket{\set{\lambda}}$ in terms of a set of $M$ parameters $\{\lambda\}$ known as {\em rapidities}. They can be calculated from the Bethe equations, a set of $M$ coupled equations
\begin{equation}
\label{eqBethe}
\left[ \frac{\phi_1(\lambda_j)} {\phi_{-1}(\lambda_j)} \right]^N 
= \prod_{k\neq j}^{1\ldots M} \frac {\phi_2(\lambda_j -\lambda_k)} {\phi_{-2}(\lambda_j -\lambda_k)}
~,
\end{equation}
with
\begin{align}
\phi_n(\lambda) &= \sinh(\lambda + n i \zeta/2) \quad\text{with $\zeta = \arccos \Delta$}
&&\quad\text{for XXZ,}
\\
\phi_n(\lambda) &= \lambda + n i/2
&&\quad\text{for XXX.} 
\end{align}
In general, equations for the XXX case $\Delta \goesto 1$ follow from those for the XXZ case by taking the limit $\zeta \goesto 0$ with $\gamma := \lambda/\zeta$ kept finite. 

To calculate the many solutions of these equations, one makes them explicit by taking the logarithm of these equations, thus introducing a set of $M$ quantum numbers $\{\tilde I\}$, which are integer for odd $M$ and half-integer for even $M$. 
\begin{align}
\label{eqLogBethe}
\theta_1^+(\lambda_j) - \frac{1}{N} \sum_{k=1}^{M} \theta_2^+ (\lambda_j-\lambda_k) &= \frac{\pi \tilde I_j}{N} 
~,
\end{align}
where
\begin{align}
\theta_n^v(\lambda) &= 2 v \arctan \left[ \tanh\lambda  / (\tan n \zeta/2)^{v} \right] &\quad\text{for XXZ; } \\
\theta_n(\lambda) &= 2 \arctan \left[ 2\lambda / n \right] &\quad\text{for XXX. } \\
\end{align}
There is an exclusion principle at work: since wave functions vanish identically for coinciding rapidities, no two rapidities must be equal. 
Because of this, for real rapidities, all quantum numbers $\tilde I_j$ are distinct. In that case, these equations can easily be solved up to high numerical precision by iteration.

\subsection{The string hypothesis and its limitations}
However, roots of Bethe's equations need not be real numbers. Already in Bethe's original paper \cite{Bethe1931} it is conjectured that complex solutions will form structures (later dubbed {\em strings}) of roots with equal real parts and imaginary parts equally spaced and distributed symmetrically around the real axis or, in the XXZ case only, the axis $i\pi/2$, as follows:
\begin{equation}
\label{eqString}
\lambda^{j}_{\alpha a} = \lambda^j_\alpha + \thalf i \zeta (n_j+1-2a) + \tquar i \pi (1-v_j) + i\delta^j_{\alpha a} \quad\text{for $a \in\set{1,\ldots,n_j}$}
~,
\end{equation}
where $j \in \{1\ldots N_s\}$ denotes the string type, $\alpha \in \{1\ldots M_k\}$ enumerates all strings of type $k$. $n_j$ is the length and $v_j \in\set{1,-1} $ the {\em parity} of a string. The string centre $\lambda^j_\alpha$ always lies on the real axis. 
In this definition we allowed for a deviation $\delta^j_{\alpha a}$; the string hypothesis is the conjecture that complex solutions of the Bethe equations take this form with $\delta^j_{\alpha a} = O(e^{-c N}) $ with $c>0$. 

For $\Delta=1$ all string lengths are allowed and have positive parity; yet, in general, what lengths are allowed and what parities are associated to them depends on the anisotropy in a rather involved way \cite{Takahashi1972, Hida1981, Fowler1981, Takahashi1999}:
if we call the number of string types $N_s$, then, for $j \in \set{1\ldots N_s}$, we have lengths and parities given by
\begin{align}
n_j &= y_{i-1} + (j-m_i)y_i
&
v_j &= (-1)^{\floor{(n_j-1)\zeta/\pi}}
\end{align}
where $N_s = m_l +1$, $i$ is chosen such that $m_i \leq j < m_{i+1}$, and the series $(y)$, $(m)$ are found from
\begin{align}
	y_{-1}&=0 ~,
	&
	y_0 &=1 ~,
	&
	y_i &= y_{i-2}+\nu_i y_{i-1}  ~,
	&&
\notag\\
	&&
	m_0 &=0 ~,
	&
	m_i &= \sum_{k=1}^i \nu_k  &&\quad\text{for $i \in\set{1,\ldots,l}$}
~,
\end{align}
in which $\nu_i \geq 1$ are the partial quotients of $\zeta/\pi$ given by
\begin{align}
\frac{\zeta}{\pi} =: 
	\cfrac{1}{\nu_1 + \cfrac{1}{\nu_2 + \cdots}}
\end{align}
where we terminate the continued fraction after a suitable number $l$ of partial quotients. Note that $\nu_l \geq 2$ to avoid an ambiguity in this definition.
In the presence of strings of length two or higher, indeterminacies of the type $\delta/\delta$ pop up in the Bethe equations. These, however, can be cancelled by taking a product of the Bethe equations for all roots constituting a string, leading to what is known as the Bethe-Takahashi equations; in logarithmic form, they are
\begin{equation}
\label{eqLogBetheTakahashi}
N \theta_{n_j}^{v_j}(\lambda^j_\alpha) - \sum_{k=1}^{N_s} \sum_{\beta=1}^{M_k} \Theta_{jk}(\lambda^j_\alpha - \lambda^k_\beta) = 2\pi I^j_\alpha
~,
\end{equation}
with 
\begin{equation}
\Theta_{jk} (\lambda) = (1-\delta_{n_j n_k}) \theta_{\abs{n_j-n_k}}^{v_j v_k} (\lambda) + 2\theta_{\abs{n_j-n_k}+2}^{v_j v_k}(\lambda) + \cdots + 2\theta_{n_j+n_k-2}^{v_j v_k}(\lambda)+ \theta_{n_j+n_k}^{v_j v_k}(\lambda)
\end{equation}
These equations give the string centres $\lambda^j_\alpha$ in terms of a new set of mutually distinct quantum numbers $I^j_\alpha$.  Note that the $M_k$ must satisfy $\sum_{k=1}^{N_s} n_k M_k = M$.

The energies of the states thus defined are
\begin{align}
E &= -J \sum_{j\alpha} \frac{ \sin\zeta \sin n_j\zeta}{v_j \cosh 2\lambda^j_\alpha - \cos n_j\alpha} - h\left(\thalf N - M \right)&&\text{for XXZ;}
\\
E &= -J \sum_{j\alpha} \frac{ 2 }{(2\lambda^j_\alpha)^2 + 1} - h\left(\thalf N - M \right)&&\text{for XXX.}
\end{align}
and their momenta are
\begin{align}
q &= \pi\sum_j \delta_{v_j,+1} M_j +  \frac{2\pi}{N} \sum_{j\alpha} I^j_\alpha \quad\text{(mod $2\pi$)}
~.
\end{align}

It is known \cite{Bethe1931, Essler1992, Ilakovac1999, Fujita2003} that the string hypothesis is not generally valid. However, after finding a numerical solution to the log-Bethe-Takahashi equations \eqref{eqLogBetheTakahashi}, one can explicitly check whether or not it is a solution of the original Bethe equations \eqref{eqBethe} to sufficient accuracy, by calculating the deviation of each string root. To first order, this deviation is given by
\begin{equation}
2 i \delta^j_{\alpha\,a} = 
	- D^j_{\alpha\,1} \cdots D^j_{\alpha\,a} \left[ 1 + D^j_{\alpha\,a+1} \left[1 + D^j_{\alpha\,a+2} 
	\left[ \cdots \left[ 1 + D^j_{\alpha\,n-a} \right] \cdots \right]\right]\right] 
\end{equation}
where
\begin{align}
D^j_{\alpha\,a} &:= 
	- \frac{\phi_{ 2(n_j - a) }(0) \phi_{2(n_j - a + 1)}(0) }{\phi_{2a}(0)  \phi_{2(a-1)}(0)  } 
	\left[ \frac{\phi_{-1}(\lambda^j_{\alpha\,a})}{\phi_1(\lambda^j_{\alpha\,a})}  \right]^{N} 
	\prod_{(k,\beta) \neq (j,\alpha)} \frac{ \phi_2 (\lambda^j_{\alpha\,a} - \lambda^k_{\beta\,b}) }{ \phi_{-2}( \lambda^j_{\alpha\,a} - \lambda^k_{\beta\,b}) }
~,
\end{align}
in which any factor of zero appearing in numerator or denominator is to be left out.

There are a few classes of string solutions which we can discard without performing any calculation: no even-length string with rapidity zero can be a solution of the Bethe equations, as the left-hand side becomes non-finite; furthermore, if there is more than one odd-length string of the same parity with rapidity zero, the exclusion principle is violated. Rapidities at zero can easily be found: if the quantum numbers are spaced symmetrically around zero, a zero quantum number implies a zero rapidity. The wave functions in these cases must be obtained by a limiting procedure on the Bethe wave functions, which we do not carry out here.

\subsection{Determinant expressions}
Although Bethe eigenstates are hard to work with, exact expressions exist for the norm \cite{Korepin1982, Gaudin1983} and overlap \cite{Slavnov1989} of states and for form factors \cite{Kitanine1999, Kitanine2000} (i.e. matrix elements of local spin operators).
By inserting a complete set of eigenstates, the structure factor can be expressed as a sum over squares of form factors of the $M$-particle ground state $\ket{\text{G}_M}$ and these eigenstates,
\begin{align}
S^{a\bar a}(q,\omega) 
	&:= \frac{1}{N}\sum_{j,j'=1}^{N} e^{iq(j-j')} \int_{-\infty}^\infty dt~ e^{i\omega t}\expect{ S^a_j(t)\,S^{\bar a}_{j'}(0) }\sub{c} 
\notag\\
	&\quad= 2\pi \sum_{\alpha \neq \text{G}_M} \abs{ \bra{ \text{G}_M } S^a_q \ket{ \alpha } }^2 \delta(\omega - \omega_\alpha)
\end{align}
The ground state itself is excluded from the sum as we calculate the connected correlator $\expect {S^a\,S^{\bar a} }\sub{c} := \expect {S^a\,S^{\bar a}} - \expect {S^a}\expect{S^{\bar a}}$.
Note that for the form factors to be non-zero $\alpha$ must be in the right subspace: for the $S^z$ form factor, with equal number $M$ of down spins as the ground state; for $S^-$, the space with $(M-1)$ down spins. 
The form factor can then be calculated using a determinant expression. For the two-particle contribution at a fixed momentum, such methods were used in \cite{Biegel2002, Biegel2003}.

We shall calculate the full contribution coming from multi-particle intermediate states over the full Brillouin zone. In terms of the string rapidities, the determinant expressions are \cite{Caux2005b}
\begin{align}
\label{eqDeterminantSz}
\abs{ \bra{ \set{\mu} } S^z_q \ket{ \set{\lambda} } }^2 &= 
	\tquar N \delta_{q,q_\lambda-q_\mu} \prod_{j}^{1\ldots M} \abs{ \frac{ \phi_{-1} (\mu_j) }{ \phi_{-1} (\lambda_j) } }^2
\times\notag\\&\quad\times 
\prod_{j\neq k}^{1\ldots M} \abs{ \phi_2 (\mu_j - \mu_k) }^{-1}
	\prod_{\substack{j\neq k \\ \phi_2(\lambda_j - \lambda_k)\neq 0}}^{1\ldots M} \abs{ \phi_2 (\lambda_j - \lambda_k) }^{-1}
\times\notag\\&\quad\times 
	\frac{ \abs{ \det \left[ \mx{H} ( \set{\mu}, \set{\lambda} ) - 2 \mx{P} ( \set{\mu},\set{\lambda} ) \right] }^2 	} 
	{ \norm{\set{\mu}} \norm{\set{\lambda} }  } 
\end{align}
and
\begin{align}
\label{eqDeterminantSm}
\abs{ \bra{ \set{\mu} } S^-_q \ket{ \set{\lambda} } }^2 &= 
	\tquar N \delta_{q,q_\lambda-q_\mu} \abs{\phi_2(0)} \frac{ \prod_{j}^{1\ldots M} \abs{  \phi_{-1} (\mu_j)}^2 }{ \prod_{j}^{1\ldots M-1} \abs{ \phi_{-1} (\lambda_j) }^2 }
\times\notag\\&\quad\times 
\prod_{j\neq k}^{1\ldots M} \abs{ \phi_2 (\mu_j - \mu_k) }^{-1}
	\prod_{\substack{j\neq k \\ \phi_2(\lambda_j - \lambda_k)\neq 0}}^{1\ldots M-1} \abs{ \phi_2 (\lambda_j - \lambda_k) }^{-1}
\times\notag\\&\quad\times 
	\frac{ \abs{ \det \left[ \mx{H}^- ( \set{\mu}, \set{\lambda} ) \right] }^2   } 
{ \norm{\set{\mu}} \norm{\set{\lambda} }  } 
~,
\end{align}
with the matrices $\mx{H}$, $\mx{H}^-$, and $\mx{P}$ given by 
\begin{align}
H^-_{an} = H_{ab} &= K_{ab}
\quad
P_{ab} = 0 
\quad\quad
\text{for $b\in \set{1\ldots n-1}$}
\notag\\
H^-_{an} = H_{an} &= 
	N \sum_{j=0}^n \frac{G_j G_{j+1}} {F_j F_{j+1}} 
	\left( [\delta_{j,0} + \delta_{j,n} -1 ] \frac{d}{d\mu_a} K_{aj} + [\delta_{j,0} + \delta_{j,n}] K_{aj} \right)
\\
P_{an} &=
	\frac{N}{\phi^2_0 (\mu_a) - \phi^2_1(0)} \sum_{j=1}^n \frac{G_j}{F_j}
\notag
\end{align}
with
\begin{align}
F_b &:= \prod_{\substack{k \neq b}} \phi_0 (\lambda_k - \lambda_b)
&
G_b &:= \prod_{k} \phi_0 (\mu_k - \lambda_b)
\\
K_{ab} &:= \left[ \phi_0(\mu_a - \lambda_b) \phi_2(\mu_a-\lambda_b) \right]^{-1}
&
N &:= F_0 F_1 G_n^{-1}\prod_{j=2}^n G_j
\notag
\end{align}
where notations such as $H_{ab}$ are a shorthand for $H_{j\alpha a, k\beta b}$ and the values $\lambda^j_{\alpha 0}$, $\lambda^j_{\alpha, n_j+1}$ are defined by extension of equation \eqref{eqString}. It should be noted that although the definitions of elements of $\mx{H}$ and $\mx{H}^-$are the same, when $\mx{H}^-$ is used the number of rapidities $\lambda$ is one less than the number of $\mu$, and therefore we have not yet defined all of its columns. The final column of $H^-$ is given by
\begin{align}
	H^-_{aM} &= \frac{1}{\phi^2_0 (\mu_a) - \phi^2_1(0)} ~.
\end{align}
The norm \cite{Korepin1982, Gaudin1983} is given by $\norm{\set{\lambda} } := \bracket{\set{\lambda}}{\set{\lambda}} = \det \mx \Phi(\set{\lambda})$ with the reduced Gaudin matrix defined as
\begin{align}
\Phi^{j\alpha, j\alpha} &:= 
	\left[ N \frac{d}{d\lambda^j_{\alpha}} \theta_j(\lambda^j_{\alpha})
	- \sum_{\substack{l\gamma \\ (l,\gamma)\neq(j,\alpha) }} \frac{d}{d\lambda^j_{\alpha}} \Theta_j(\lambda^j_{\alpha}-\lambda^l_\gamma) \right] 
~,
 \notag\\
\Phi^{j\alpha, k\beta} &:= 	\frac{d}{d\lambda^j_{\alpha}} \Theta_j(\lambda^j_{\alpha}-\lambda^l_\gamma)
\quad\text{for $(j,\alpha)\neq(k,\beta)$}
\end{align}

\subsection{Enumeration of states}
With $2^N$ states in the Hilbert space, the form factor expansion by itself does not suffice to calculate the structure factor for anything but very small systems. What saves the day is that the number of states over which we have to sum can be vastly reduced \cite{Karbach2000, Karbach2002, Sato2004, Caux2005a}. 

First of all, as we already noted, we only need a single $M$ subspace.
Secondly, we can divide this subspace further into classes of states, which we shall call {\em bases}, distinguished by the number and character of excitations they contain. For instance, there is always a base containing only real rapidities, which we could denote $(\cdot)$. Then, for the XXZ chain, the first different type of particle is a rapidity with negative parity; if we have one of them we have base $(\cdot,1)$. A base with three negative-parity rapidities and a two-string would then be $(\cdot,3,1)$, and so on. 

For each such base, we can determine bounds for the quantum numbers $\set{I}$. We do this by letting the largest rapidity in the log-Bethe-Takahashi equation \eqref{eqLogBetheTakahashi} go to infinity; since the order of rapidities equals the order of quantum numbers, all quantum numbers must be lower than or equal to the one that leads to rapidity infinity. Allowed quantum numbers $I$ for string type $j$ thus satisfy 
\begin{equation}
2\abs{I^j} \leq 2I^j\sub{bound} := \frac{1}{\pi} \abs{ N \theta_j(\infty) -\sum_{k=1}^{N_s} (M_k - \delta_{jk})\, \Theta_{jk}(\infty)  } 
\end{equation}
where it is understood that all $I^j$ are integer for odd $M_j$, half-integer for even $M_j$. Let us call the highest allowed quantum number $I^j_\infty$; then, the number of available positions for string type $j$ is $2I^j_\infty + 1$. 
If the equality $I^j_\infty = I^j\sub{bound}$ is satisfied, as happens in the XXX case, we have a rapidity that is formally infinite. These rapidities merit special attention which we shall bestow on them later on.

First, let us turn our attention to the lowest string type: the real rapidities. Since the ground state forms a Fermi interval in terms of the quantum numbers, any occupied position outside this interval is a particle and any empty position is a hole. As for the other string types, every occupied position there counts as a particle.

It is observed that the sum of form factors in a base decreases rapidly with the total number of particles of each type (particles, holes, and higher strings). Therefore, we order our states along these lines and sum over their form factors progressively to get an approximate value for the structure factor. We can check the accuracy of this approximation by comparing the total sum to the sum rules
\begin{align}
2\pi \sum_q \sum_{\alpha \neq \text{G}_M} \abs{ \bra{ \text{G}_M } S^z_q \ket{ \alpha } }^2 
	&= \tquar - \expect{S^z}^2 
	= \tquar\left[1-\left(1-2\frac{M}{N}\right)^2\right] ~\text{, }
\\
2\pi \sum_q \sum_{\alpha \neq \text{G}_M} \abs{ \bra{ \text{G}_M } S^-_q \ket{ \alpha } }^2 
	&= \thalf - \expect{S^z} 
	= \frac{M}{N} ~.
\end{align}

In the isotropic case, at zero field, we can also check against the sum rule for the first frequency moment, 
\begin{align}
\sum_{\alpha \neq \text{G}_M} \abs{ \bra{ \text{G}_M } S^z_q \ket{ \alpha } }^2 = \tfrac{2}{3}(1 - \cos q)(\tfrac{1}{4} - \ln 2) 
~,
\end{align}
giving an extra check on each slice of fixed momentum $q$.

\subsection{Infinite rapidities}
In the isotropic model, from a given solution of the Bethe-Takahashi equations at $M$ down spins, one can generate a cascade of solutions at $M+1, M+2, \ldots N/2$ by applying the descending operator $S^-_0$, without changing the existing rapidities. The action of the descending operator is equivalent to adding an extra rapidity at infinity:
\begin{equation}
S^-_0 \ket{ \set{\gamma} } = \ket{\set{\gamma, \pm\infty}} 
\end{equation}
The momentum of a state is left unchanged by this modification and the energy is raised by $h\Delta M = h$ in the presence of an external magnetic field $h$. 
The form factors of such a state are also easy to calculate: we have the relations
\begin{align}
\abs{ \bra{ \set{\mu} } S^-_q \ket{ \set{\gamma, \pm \infty} } }^2 &= 0
\\
\abs{ \bra{ \set{\mu} } S^z_q \ket{ \set{\gamma, \pm \infty} } }^2 &= 
	\frac{\abs{ \bra{ \set{\mu} } S^-_q \ket{ \set{\gamma} } }^2}
{N-2M_{\set{\gamma}}}
\end{align}
as can be shown explicitly from the determinant expressions \eqref{eqDeterminantSz}, \eqref{eqDeterminantSm} or by application of the Wigner-Eckart theorem, noting the rotational symmetry of the Hamiltonian and the vector nature of the spin operator \cite{Mueller1981}.

\section{Results}
\begin{figure*}
\label{figSmp}
\includegraphics[width=13.5cm]{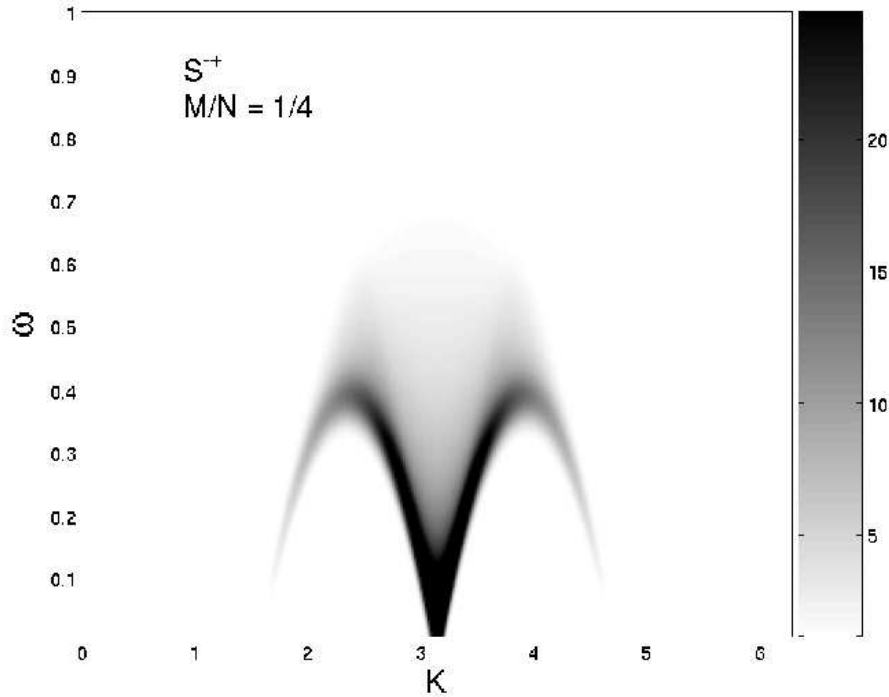}
\caption{Transverse correlation function $S^{-+}(K, \omega)$ at $\Delta=1$, $N=320$, and $M=N/4=80$. The energy $\omega$ is shown in units of the coupling $J$. In this calculation, the sum rule was satisfied to $98.6\%$. }
\end{figure*}

As an example of the quality of results we can achieve with this method, in Figure \ref{figSmp} we show the transverse correlation function for an isotropic chain of $N=320$ sites at $M=N/4$. More results for anisotropic chains can be found in \cite{Caux2005b}. 

\section*{Acknowledgements}
R.H. would like to thank I.I.A.S.S. for their hospitality during the Xth Training Course in the Physics of Correlated Electron Systems and High-T$_c$ Superconductors in Vietri sul Mare.

\bibliographystyle{aipproc}   

\begin{thebibliography}{9}
\bibitem{Bethe1931}					H.~Bethe, \emph{Z. Phys}, \textbf{71} 205--226 (1931).
\bibitem{Orbach1958}			R.~Orbach, \emph{Phys. Rev.} \textbf{112}, 309--316 (1958).
\bibitem{Takahashi1971}			M.~Takahashi, \emph{Prog. Theor. Phys.}, \textbf{46} 401--415 (1971).
\bibitem{Takahashi1972}			M.~Takahashi, M.~Suzuki, \emph{Prog. Theor. Phys.}, \textbf{48} 2187--2209 (1972).
\bibitem{Hida1981}		K.~Hida, \emph{Phys. Lett. A}, \textbf{84} 338--340 (1981).
\bibitem{Fowler1981}	M.~Fowler, X.~Zotos, \emph{Phys. Rev. B} \textbf{24} 2634--2639 (1981).
\bibitem{Takahashi1999}	M.~Takahashi, \emph{Thermodynamics of One-Dimensional Solvable Models}, Cambridge University Press, Cambridge, 1999, pp. 133--138.
\bibitem{Korepin1982} 	V.~E.~Korepin, \emph{Commun. Math. Phys.}, \textbf{86} 391--418 (1982).
\bibitem{Gaudin1983}	M.~Gaudin, \emph{La fonction d'onde de Bethe}, Masson, Paris, 1983, pp. 67--68.
\bibitem{Slavnov1989}				N.~A.~Slavnov, \emph{Theor. Math. Phys.} \textbf{79} 502--509 (1989).
\bibitem{Kitanine1999}			N.~Kitanine, J.~M.~Maillet, V.~Terras, \emph{Nucl. Phys. B.} \textbf {554} 647--678 (1999).
\bibitem{Kitanine2000} 			N.~Kitanine, J.~M.~Maillet, V.~Terras, \emph{Nucl. Phys. B.} \textbf {567} 554--584 (2000).
\bibitem{Mueller1981}				G.~M\"{u}ller, H.~Thomas, H.~Beck, J.~C.~Bonner, \emph{Phys. Rev. B} \textbf{24} 1429--1467 (1981).
\bibitem{Biegel2002}		D.~Biegel, M.~Karbach, G.~M\"{u}ller,	\emph{Europhys. Lett.} \textbf{59} 882--888 (2002).
\bibitem{Biegel2003}		D.~Biegel, M.~Karbach, G.~M\"{u}ller, \emph{J. Phys. A: Math. Gen.} \textbf{36} 5361--5368 (2003).
\bibitem{Essler1992}				F.~H.~L.~Essler, V.~E.~Korepin, K.~Schoutens, \emph{J. Phys. A: Math. Gen} \textbf{25} 4115--4126 (1992).
\bibitem{Ilakovac1999} 				A.~Ilakovac, M.~Kolanovi\'{c}, S.~Pallua, P.~Prester, \emph{Phys. Rev. B} \textbf{60} 7271--7277 (1999).
\bibitem{Fujita2003}				T.~Fujita, T.~Kobayashi, H.~Takahashi, \emph{J. Phys. A: Math. Gen.} \textbf{36} 1553--1564 (2003).
\bibitem{Karbach2000}				M.~Karbach, and G.~M\"{u}ller, \emph{Phys. Rev. B} \textbf{62}, 14871--14879 (2000).
\bibitem{Karbach2002}				M.~Karbach, D.~Biegel, G.~M\"{u}ller, \emph{Phys. Rev. B} \textbf{66}, 054405 (2002).
\bibitem{Sato2004}					J.~Sato, M.~Shiroishi, M.~Takahashi, \emph{J. Phys. Soc. Japan} \textbf{73} 2008--3014 (2004).
\bibitem{Caux2005a}					J.-S.~Caux, J.~M.~Maillet, \emph{Phys. Rev. Lett} \textbf{95} 077201 (2005).
\bibitem{Caux2005b}					J.-S.~Caux, R.~Hagemans, J.~M.~Maillet, \emph{J. Stat. Mech.} P09003 (2005).
\end{thebibliography}

\end{document}